\documentclass[sigconf,natbib]{acmart}
\usepackage{array}
\usepackage{graphicx}
\usepackage{amsmath}

\usepackage[table]{xcolor}
\usepackage{xfp}   % floating point comparisons

\def\figscale{.51}
\newcommand{\boldsec}[1]{\ensuremath{\texorpdfstring{\boldsymbol{#1}}{#1}}}

\newcommand{\yes}{\ensuremath{\bulletS}}
\newcommand{\altrow}{\rowcolor{gray!10}}

\copyrightyear{2026}
\acmYear{2026}
\setcopyright{cc}
\setcctype{by}
\acmConference[SIGIR '26]{Proceedings of the 49th International ACM SIGIR Conference on Research and Development in Information Retrieval}{July 20--24, 2026}{Melbourne, VIC, Australia}
\acmBooktitle{Proceedings of the 49th International ACM SIGIR Conference on Research and Development in Information Retrieval (SIGIR '26), July 20--24, 2026, Melbourne, VIC, Australia}
\acmDOI{10.1145/3805712.3808540}
\acmISBN{979-8-4007-2599-9/2026/07}

\begin{document}

\title[Stop Using the Wilcoxon Test: Myth, Misconception and Misuse in IR Research]{Stop Using the Wilcoxon Test:\\Myth, Misconception and Misuse in IR Research}

\author{Juli\'an Urbano}
\orcid{0000-0003-2933-1949}
\affiliation{
	\institution{Delft University of Technology}
	\city{Delft}\country{The Netherlands}
}
\email{j.urbano@tudelft.nl}

\sloppy
\begin{abstract}
In benchmarking of Information Retrieval systems, the Wilcoxon signed-rank test is often treated as a safer alternative to the $t$-test. This belief is fueled by textbooks and recommendations that portray Wilcoxon as the proper non-parametric alternative because metric scores are not normally distributed. We argue that this narrative is misleading and harmful. A careful review of Statistics textbooks reveals inconsistencies and omissions in how the assumptions underlying these tests are presented, fostering confusion that has propagated into IR research. As a result, Wilcoxon has been routinely misapplied for decades, creating a false sense of safety against a threat that was never there to begin with, while introducing another one so severe that it virtually guarantees the test will break down and mislead researchers. Through a combination of systematic literature review, analysis and empirical demonstrations with TREC data, we show how and why the Wilcoxon test easily loses control of its Type~I error rate in IR settings. We conclude that the continued use of Wilcoxon in IR evaluation is unjustified and that abandoning it would improve the methodological soundness of our field.\looseness=-1
\end{abstract}

\maketitle

%%%%%%%%%%%%%%%%%%%%%%%%%%%%%%%%%%%%%%%%%%%%%%%%%%%%%%%%%%%%%%%%%%%%%%%%%%%%%%%%%%%%%%%%%%%%%%%%%%%%
\section{Introduction}\label{sec:intro}
%%%%%%%%%%%%%%%%%%%%%%%%%%%%%%%%%%%%%%%%%%%%%%%%%%%%%%%%%%%%%%%%%%%%%%%%%%%%%%%%%%%%%%%%%%%%%%%%%%%%

In Information Retrieval (IR) evaluation, system effectiveness is usually measured over the fixed set of topics in a test collection or dataset. Because topic difficulty varies, observed differences between two systems, say \textsf{X} and \textsf{Y}, inevitably reflect both genuine performance gaps and random variation due to topic sampling. Statistical testing is typically used to separate signal from noise: to ask whether the apparent improvement of \textsf{X} over \textsf{Y} is more than a coincidence due to the selection of topics in the test collection.

Formally, for each topic $i=1,\dots,n$ we record scores $X_i$ and $Y_i$ according to some metric like NDCG, and the null hypothesis of interest is that the systems have equal mean effectiveness in the population of topics, $H_0:\mu_X=\mu_Y$. Since both systems are tested on the same topics, this is called a \emph{paired two-sample problem}, where dependence across topics (difficulty) is accounted for. This leaves us with two classical procedures that dominate the Statistics literature and IR practice, usually described as follows:
\begin{enumerate}
	\item Paired Student $t$-test: a parametric test assuming that scores follow a normal distribution.% It relies on the mean and variance of the observed scores.
	\item Wilcoxon signed-rank test: a non-parametric alternative when the normality assumption is violated.% It ignores raw magnitudes and instead evaluates whether positive differences tend to outweigh negative ones.
\end{enumerate}
\noindent It is worth noting that once we define the per-topic differences $D_i=X_i-Y_i$, this problem is equivalent to testing the one-sample hypothesis $H_0:\mu_D=0$. As we will show in Section~\ref{ssec:t}, this alternative view simplifies definitions and discussion.

Over the decades, IR research has repeatedly asked which test is ``best'' for system benchmarking. A first wave of papers discussed mostly theoretical arguments~\cite{rijsbergen1979information,hull1993using,wilbur1994non}. % sign and bootstrap
A second wave took an empirical angle, using resampling and random topic set splits~\cite{savoy1997statistical,zobel1998how,voorhees2002effect,sanderson2005information,sakai2006evaluating,smucker2007comparison,cormack2007validity,smucker2009agreement,carterette2012multiple,urbano2013comparison}. %resampling, t
A third wave is represented by simulation studies that recently calculated actual Type~I and II errors for IR-like data,\footnote{As a reminder, a Type~I error occurs when the null hypothesis is true but it is rejected (claiming a difference when there is none, i.e., a false positive), and a Type~II error occurs when the null is not true and yet it is not rejected (missing a difference when there is one, i.e., a false negative).} but reached opposite conclusions: \cite{urbano2018stochastic,urbano2019statistical,urbano2021how} recommend the $t$-test and discourage the Wilcoxon test, while \cite{parapar2020using,parapar2021testing} actually recommend the latter. In practice, surveys of venues like SIGIR and ECIR show that about 65\% of papers use the $t$-test and 25\% Wilcoxon, with other methods being a minority~\cite{sakai2016statistical,carterette2017is}.
	
In this paper, we argue that the Wilcoxon test should no longer be used in IR evaluation, not because it is sub-optimal, but because it is actively \emph{harmful}. Unlike earlier empirical or simulation-based work, we take a different route and present a systematic review of 25 Statistics textbooks. This reveals the root of a problem: the very sources we rely upon to shape our collective understanding perpetuate a confusing and misleading dichotomy between parametric and non-parametric methods. We frame the issue in terms of \emph{myth}, \emph{misconception}, and \emph{misuse}. The myth is that non-parametric methods are inherently safer when parametric assumptions fail. The misconception is that the $t$-test requires normally distributed scores, when in fact it only requires their mean to be approximately normal. The misuse is the reliance on the Wilcoxon test as a quick remedy when normality is questioned, because its own assumption of \emph{symmetry} is actually far stronger, yet almost always ignored. Through analysis and simulation, we show that this oversight makes the Wilcoxon test fail catastrophically for common IR data, whereas the $t$-test remains robust. In short, the supposed ``safe'' alternative actually turns out to be the riskier choice, and continuing its use only deepens the confusion and undermines the credibility of our findings.\looseness=-1
\footnote{Data and code are available at \url{https://github.com/julian-urbano/sigir2026-wilcoxon}.}

%All the results in this paper can be fully reproduced with data and code available online.\footnote{\url{https://github.com/julian-urbano/sigir2026-wilcoxon}.}

%%%%%%%%%%%%%%%%%%%%%%%%%%%%%%%%%%%%%%%%%%%%%%%%%%%%%%%%%%%%%%%%%%%%%%%%%%%%%%%%%%%%%%%%%%%%%%%%%%%%
\section{Systematic Review of the Statistics Literature}\label{sec:stats-lit}
%%%%%%%%%%%%%%%%%%%%%%%%%%%%%%%%%%%%%%%%%%%%%%%%%%%%%%%%%%%%%%%%%%%%%%%%%%%%%%%%%%%%%%%%%%%%%%%%%%%%

In order to understand the confusion around statistical testing, we review a selection of textbooks to investigate how the Statistics literature describes these methods. In particular, we selected a total of 25 textbooks, categorized as follows:
\begin{itemize}
	\item 11 books on Statistics in general \cite{box2005statistics,cetinkaya2024introduction,dekking2005modern,kanji2006statistical,lehmann2022testing,mendenhall2018introduction,ott2015introduction,rice2007mathematical,sheskin2000handbook,valiela2001doing,weerahandi2003exact}.
	\item 6 books specifically on Non-parametric Statistics \cite{conover1999practical,hettmansperger1984statistical,hollander1973nonparametric,neuhauser2012nonparametric,pratt1981concepts,siegel1956nonparametric,wasserman2006all}.
	\item 8 books on Statistics for particular fields such as Behavioral Sciences \cite{privitera2014statistics}, Biology \cite{sokal1995biometry}, Business and Economics \cite{anderson2019statistics}, Computer Science \cite{baron2014probability}, Education \cite{cohen2018research}, Engineering \cite{montgomery2014applied}, Environmental Science \cite{manly2008statistics}, and Psychology \cite{aron2013statistics}.
\end{itemize}

We examined and annotated all these sources with respect to how they discuss recommendations, non-parametric methods, the $t$-test, and the Wilcoxon test.

\subsection{Recommendations}\label{ssec:recs}
%%%%%%%%%%%%%%%%%%%%%%%%%%%%%%%%%%%%%%%%%%%%%%%%%%%%%%%%%%%%%%%%%%%%%%%%%%%%%%%%%%%%%%%%%%%%%%%%%%%%

When researchers seek guidance on which statistical test to use, many sources offer quick roadmaps with questions like \textsl{``how many groups do you have?''} or \textsl{``are your data normally distributed?''} to guide them. Of the 25 books we reviewed, 10 provided such shortcuts in the form of tables or decision trees (see Table~\ref{tab:roadmap}).

The first noteworthy observation is the diversity of descriptions about the object of the test leading to the paired-sample Wilcoxon test, ranging from vague statements like ``differences between groups,'' to precise ones such as ``location, median,'' or confusing ones such as ``the ordering of data in two dependent populations.'' As the table shows, most sources contrast parametric and non-parametric methods, and/or link test choice to the level of measurement of the data. In the latter cases, Wilcoxon is recommended for ordinal data, with the exception of \cite[pp. 252]{conover1999practical}, who argues that $D$ being ordinal implies that both $X$ and $Y$ must be interval. On the other hand, \cite[pp. 76]{siegel1956nonparametric} argues for $X$ and $Y$ on an ``ordered metric'' scale, which lies between ordinal and interval in strength \cite{coombs1950psychological}.\looseness=-1

The diversity and vagueness in Table~\ref{tab:roadmap} suggest that a proper understanding of the underlying principles is essential. In the worst case, a researcher may simply pick ``whatever seems to fit my problem''. In the best case, they find quick guidance but are expected to grasp the subtleties, their implications and risks. This raises the question: how does the Statistics literature actually explain concepts such as \textsl{non-parametric methods}, and does it enable sound choices?\looseness=-1

\begin{table}[t]
	\caption{Textbooks with decision trees for selecting the appropriate statistical test. The main decision is made with respect to parametric vs. non-parametric methods, or level of measurement (for these, the table lists the level specified for the Wilcoxon test). The reported object of the test refers to the description used to lead into the paired-sample Wilcoxon test. Implicit mentions are indicated with square brackets.}\label{tab:roadmap}
	
	\centering{\footnotesize\setlength{\tabcolsep}{1mm}\begin{tabular}{|rp{47mm}|c|>{\centering\arraybackslash}p{14mm}|}
			\hline
			& & \textbf{Param vs.} & \textbf{By level,} \\
			& \textbf{Object of the test} & \textbf{Non-param} & \textbf{Wilcoxon as} \\ \hline\hline
			
			\multicolumn{4}{|c|}{\textbf{General}} \\\hline
			\altrow\cite{kanji2006statistical} & Central tendency & \yes$^\text{a}$ &  \\
			\cite{sheskin2000handbook} & Hypothesis about the ordering of data in two dependent populations &  & ordinal /\linebreak rank-order \\ 
			\altrow\cite{valiela2001doing} & Difference between two samples or groups & \yes$^\text{b}$ & ordinal / non-param$^\text{b}$ \\ \hline\hline
			
			\multicolumn{4}{|c|}{\textbf{Non-parametric}} \\\hline
			\altrow\cite{conover1999practical} & Means (medians) & [\yes] & interval \\ 
			\cite{hollander1973nonparametric} & Location, median & [\yes] &  \\ 
			\altrow\cite{siegel1956nonparametric} & -- & [\yes] & ordinal$^\text{c}$ \\ \hline\hline
			
			\multicolumn{4}{|c|}{\textbf{Field-specific}} \\\hline
			\altrow\cite{aron2013statistics} & -- & \yes$^\text{d}$ &  \\ 
			\cite{cohen2018research} & Difference & \yes & ordinal \\ 
			\altrow\cite{montgomery2014applied} & Difference in means from two normal distributions in a paired analysis & \yes$^\text{e}$ &  \\ 
			\cite{privitera2014statistics} & Differences between groups & \yes & ordinal \\ \hline
			
			\multicolumn{4}{l}{$^\text{a}$ The dichotomy is actually parametric vs. distribution-free.}\\
			\multicolumn{4}{l}{$^\text{b}$ The dichotomy is actually measurement (parametric) vs. ordinal (non-parametric).}\\
			\multicolumn{4}{l}{$^\text{c}$ $D$ must also be ordinal, not just $X$ and $Y$.}\\
			\multicolumn{4}{l}{$^\text{d}$ The dichotomy is actually parametric vs. rank-order.}\\
			\multicolumn{4}{l}{$^\text{e}$ Only lists non-parametric tests for the unpaired case.}
	\end{tabular}}
\end{table}

\subsection{The Myth: Non-parametric Safety}\label{ssec:nonparam}
%%%%%%%%%%%%%%%%%%%%%%%%%%%%%%%%%%%%%%%%%%%%%%%%%%%%%%%%%%%%%%%%%%%%%%%%%%%%%%%%%%%%%%%%%%%%%%%%%%%%

\begin{table*}[t]
\caption{Description of \textsl{non-parametric methods} in Statistics textbooks. They may mention few(er) or no assumptions in general (G), about the distribution (D), about its functional form (F) or about its parameters (P). Some say they are also called \textsl{distribution-free}, and some give conflicting definitions. Some key excerpts are slightly rewritten due to space constraints.}\label{tab:nonparam}

\centering{\footnotesize\setlength{\tabcolsep}{1mm}\begin{tabular}{|rl|cc|cc|}
\hline
& & \multicolumn{4}{c|}{\textbf{Assumptions}} \\\cline{3-6}
& \textbf{Key excerpt} & \textbf{None~} & \textbf{~Few(er)~} & \textbf{Dist-free} & \textbf{Conflict} \\ \hline\hline

\multicolumn{6}{|c|}{\textbf{General}} \\\hline
\altrow\cite{box2005statistics} & ``replace the normal-iid assumption for the assumption of exchangeability'' & ~ & ~ & \yes & ~ \\ 
\cite{dekking2005modern} & ``does not assume a particular parametric model'' & F & ~ & ~ & ~ \\ 
\altrow\cite{kanji2006statistical} & ``no assumption has to be made regarding the frequency distribution [...] parametric only if normal'' & D & ~ & ~ & ~ \\ 
\cite{lehmann2022testing} & ``without assuming a given functional form for the distribution, such as the normal'' & F & ~ & ~ & ~ \\ 
\altrow\cite{ott2015introduction} & ``alternative when the population distribution is non-normal'' & ~ & ~ & ~ & ~ \\
\cite{mendenhall2018introduction} & ``few or no distributional assumptions are required [...] hypotheses in terms of population distributions, not parameters
'' & D & D & ~ & ~ \\ 
\altrow\cite{rice2007mathematical} & ``do not assume that the data follow any particular distributional form'' & F & ~ & ~ & ~ \\ 
\cite{sheskin2000handbook} & ``no assumptions with regard to the population parameters that characterize the distributions [...] really not assumption free'' & P & ~ & ~ & ~ \\ 
\altrow\cite{valiela2001doing} & ``make no assumptions about distributions'' & D & ~ & ~ & ~ \\ 
\cite{weerahandi2003exact} & ``do not require the specification of the underlying distribution [...] no common agreement among statisticians'' & F & ~ & \yes & ~ \\
\altrow\cite{cetinkaya2024introduction} & -- & ~ & ~ & ~ & ~ \\  \hline\hline

\multicolumn{6}{|c|}{\textbf{Non-parametric}} \\\hline
\altrow\cite{conover1999practical} & ``do not assume a particular population probability distribution [...] valid for any population with any distribution'' & D, F & ~ & ~ & \yes$^\text{a}$ \\
\cite{hettmansperger1984statistical} & ``based on ranks'' & ~ & ~ & ~ & ~ \\ 
\altrow\cite{hollander1973nonparametric} & ``require few assumptions about the underlying populations [...] in particular, the traditional assumption that they are normal'' & ~ & D & ~ & ~ \\  
\cite{pratt1981concepts} & ``when important test properties hold even if only very general assumptions are made about the probability distributions'' & ~ & G & \yes & ~ \\ 
\altrow\cite{siegel1956nonparametric} & ``no conditions about the parameters of the population [...] assumptions are fewer and much weaker'' & F, P & G & \yes & \yes$^\text{b}$ \\\hline\hline

\multicolumn{6}{|c|}{\textbf{Field-specific}} \\\hline
\altrow\cite{anderson2019statistics} & ``without making an assumption about the specific form of the population's probability distribution'' & F & ~ & \yes & ~ \\ 
\cite{aron2013statistics} & ``no assumptions about the shape of populations [...] or about population parameters'' & F, P & ~ & \yes & \yes$^\text{c}$ \\ 
\altrow\cite{baron2014probability} & ``does not assume any particular distribution'' & F & ~ & ~ & ~ \\			
\cite{cohen2018research} & ``few or no assumptions about the distribution [...] no assumptions about how normal, even and regular the distribution'' & F & D & ~ & \yes$^\text{d}$ \\ 			
\altrow\cite{manly2008statistics} & ``often a question of whether the population has a normal distribution or not [...] parametric tests require more assumptions'' & ~ & G & ~ & ~ \\ 
\cite{montgomery2014applied} & ``does not depend on the form of the underlying distribution of the observations'' & F & ~ & \yes & ~ \\ 
\altrow\cite{neuhauser2012nonparametric} & ``neither require a specific distributional assumption nor a high level of measurement'' & F & ~ & ~ & ~ \\ 
\cite{privitera2014statistics} & ``do not require that the data in the population be normality distributed [...] data can have any type of distribution'' & D & ~ & \yes & \yes$^\text{e}$ \\ 
\altrow\cite{sokal1995biometry} & ``their null hypothesis is not concerned with specific parameters but only with the distribution of the variables'' & P & G & \yes & ~ \\  \hline

\multicolumn{6}{l}{$^\text{a}$ ``No particular distribution'' does not mean ``\emph{any} distribution''.}\\
\multicolumn{6}{l}{$^\text{b}$ Scores not distributed in a certain way [...] but \emph{still} certain assumptions.}\\
\multicolumn{6}{l}{$^\text{c}$ ``No assumption about \emph{parameters}'' of the distribution does not mean ``no assumptions about \emph{shape}''.}\\
\multicolumn{6}{l}{$^\text{d}$ First says ``\emph{few or no} assumptions'', and later says ``\emph{no} assumptions''.}\\
\multicolumn{6}{l}{$^\text{e}$ ``Non-normal distribution'' does not mean ``\emph{any} distribution''.}
\end{tabular}}
\end{table*}

Defining \textsl{non-parametric methods} precisely proves so difficult that some authors avoid the question altogether. For example, L.~Wasserman, in his popular \textsl{All of Nonparametric Statistics}~\cite{wasserman2006all}, ``did not venture to give one, no doubt I would be barraged with dissenting opinions.'' He simply says that they require ``as few assumptions as possible'' and proposes the alternative term ``infinite-dimensional.'' Similar hesitation appears across our 25 books~\cite{box2005statistics,conover1999practical,hettmansperger1984statistical,pratt1981concepts,weerahandi2003exact}, with some of them simply choosing to not give a definition at all, however vague~\cite{cetinkaya2024introduction,ott2015introduction}.

Table~\ref{tab:nonparam} summarizes how the books describe non-parametric statistics. Diversity is again the norm, with passages like ``few or no assumptions'', ``no assumptions about how normal'' or ``no assumption about distribution parameters''. We can generally classify books in two groups: 18 say that these methods do not make \emph{any} assumptions, and 7 say that they make \emph{few or fewer} assumptions. In 4 books, these assumptions are specified in general or vague terms, sometimes not even clarifying what the object of the assumption is, while in another 7 they clearly talk about assumptions regarding the distributions. In 12 books the authors explicitly talk about \emph{specific} distributions, their shape or functional forms, while 4 books narrow it down to the specific parameters of these distributions. 
This differentiation might seem nitpicky, but it is an important one. Imagine a test that only assumes that distributions are symmetric:\footnote{This is in fact an assumption of the Wilcoxon signed-rank test (see Section~\ref{ssec:wilcox}).} this is certainly not distribution-free, it does not deal with a specific distribution family or its parameters, but it definitely constrains its shape. It is therefore understandable how the statements in Table~\ref{tab:nonparam} might mislead a reader unaware of the details. In fact, because of the difficulty in properly defining the term, 5 books end up giving inconsistent statements.

Acknowledging this confusion, some authors propose alternative terms such as \textsl{exchangeability theory} methods~\cite{box2005statistics} or \textsl{assumption freer}~\cite{sheskin2000handbook}, while others prefer to emphasize their use of ranks~\cite{aron2013statistics,hettmansperger1984statistical,mendenhall2018introduction}. However, as many as 9 books explicitly call them \textsl{distribution-free}, reinforcing a false sense of generality that they do not actually possess because they, in fact, \emph{do make assumptions} about distributions. As Box et al.~\cite{box2005statistics} noted, practitioners should be forgiven for being misled by such terminology.

In practice, guidelines very often reduce the decision to whether normality holds or not, but rejecting normality is trivial in IR evaluation with simple arguments like ``metric scores are discrete'' or ``metric scores are bounded in [0,1]'' (see e.g. \cite{carterette2012multiple} for a larger discussion). Such arguments, while undeniably proving non-normality, may have catastrophic consequences as researchers would automatically lean towards non-parametric methods, not realizing that they make assumptions of their own, so much that they may actually pose more severe risks than their parametric counterparts. We illustrate this next, showing that the parametric Student $t$-test behaves just fine despite assuming normality, while the non-parametric Wilcoxon signed-rank test breaks dramatically when its own assumptions, often unheard of, are slightly violated.

\subsection{The Misconception: Student $\boldsec{t}$-test}\label{ssec:t}
%%%%%%%%%%%%%%%%%%%%%%%%%%%%%%%%%%%%%%%%%%%%%%%%%%%%%%%%%%%%%%%%%%%%%%%%%%%%%%%%%%%%%%%%%%%%%%%%%%%%

\begin{table*}[t]
\caption{Description of the $\boldsec{t}$-test and Wilcoxon test in Statistics textbooks. The first column indicates whether they describe the equivalence between the paired two-sample and one-sample problems. For $\boldsec{t}$-test: they may place the normality assumption on the scores (X), their difference (D), or the means (M); some of them discuss behavior for large $\boldsec{n}$, and some give specific warnings depending on the data. For Wilcoxon: they may mention the continuity assumption and the consequences of zeros and ties, as well as the assumption of symmetry and the consequence of actually testing the median. }\label{tab:t-wilcox}
\centering{\footnotesize\setlength{\tabcolsep}{1mm}\begin{tabular}{|r|c|ccc|ccccc|}\hline
	\multicolumn{1}{|c}{}	& \textbf{paired =} & \multicolumn{3}{c|}{\textbf{$\boldsec{t}$-test}} & \multicolumn{5}{c|}{\textbf{Wilcoxon test}} \\ 
\multicolumn{1}{|c}{} & \textbf{1-sample} & \textbf{Normality} & \textbf{Large $\boldsec{n}$} & \textbf{Warnings} & \textbf{Continuity} & \textbf{Zeros} & \textbf{Ties} & \textbf{Symmetry} & \textbf{Median}  \\ \hline\hline

\multicolumn{10}{|c|}{\textbf{General}} \\\hline
\altrow\cite{box2005statistics}  & [\yes]$^\text{a}$ & ?, [M] &  &  &  &  &  &  &   \\ 
\cite{dekking2005modern}  &  & ?$^\text{b}$ & \yes & Small $n$ &  &  &  &  &   \\ 
\altrow\cite{kanji2006statistical}  & [\yes]$^\text{a}$ & X &  &  & \yes &  &  & [\yes]$^\text{b}$ &   \\ 
\cite{lehmann2022testing}  & \yes & D & \yes & Heavy tails &  &  &  & \yes &   \\ 
\altrow\cite{mendenhall2018introduction}  & \yes & X, [D]$^\text{c}$ & \yes &  &  & \yes$^\text{d}$ & \yes$^\text{d}$ &  &   \\ 
\cite{ott2015introduction}  & \yes & D &  & Skewness &  & \yes$^\text{d}$ & \yes$^\text{d}$ & \yes & \yes  \\ 
\altrow\cite{rice2007mathematical}  & [\yes]$^\text{a}$ & D & \yes & Small $n$, non-normal &  &  & \yes & \yes &   \\ 
\cite{sheskin2000handbook}  &  & X &  &  &  &  &  & \yes & \yes  \\ 
\altrow\cite{valiela2001doing}  &  & [?]$^\text{e}$ &  &  &  &  &  &  &   \\ 
\cite{weerahandi2003exact}  & \yes & X, [D] &  &  &  & \yes$^\text{d}$ & \yes$^\text{d}$ & \yes & \yes  \\
\altrow\cite{cetinkaya2024introduction}  & \yes & M, [D] & \yes &  &  &  &  &  &   \\ \hline\hline

\multicolumn{10}{|c|}{\textbf{Non-parametric}} \\\hline
\altrow\cite{conover1999practical}  & \yes & D &  &  & & \yes$^\text{d}$ & \yes$^\text{d}$ & \yes & \yes  \\ 
\cite{hettmansperger1984statistical}  &  & ?$^\text{b}$ &  &  &  &  &  & \yes & \yes  \\ 
\altrow\cite{hollander1973nonparametric}  &  &  &  &  & \yes & \yes & \yes & \yes & \yes  \\ 
\cite{neuhauser2012nonparametric}  &  &  &  &  &  &  &  & \yes & \yes  \\ 
\altrow\cite{pratt1981concepts}  &  &  &  & & \yes & \yes & \yes & \yes & \yes  \\ 
\cite{siegel1956nonparametric}  &  & D &  & Assumptions unrealistic &  &  &  &  &   \\ \hline\hline

\multicolumn{10}{|c|}{\textbf{Field-specific}} \\\hline
\altrow\cite{anderson2019statistics}  & \yes & D & \yes & &  & \yes$^\text{d}$ & \yes$^\text{d}$ & \yes & \yes  \\ 
\cite{aron2013statistics}  & \yes & D & \yes & Skewness &  &  &  &  &   \\ 
\altrow\cite{baron2014probability}  &  & ?$^\text{b}$ &  &  & \yes & \yes & \yes & \yes & \yes  \\ 
\cite{cohen2018research}  & [\yes] & ? & \yes &  &  &  &  &  &   \\ 
\altrow\cite{manly2008statistics}  & \yes & D &  &  &  &  &  &  &   \\ 
\cite{montgomery2014applied}  & \yes & D & \yes & Skewness, not unimodal & \yes &  & \yes & \yes & \yes  \\ 
\altrow\cite{privitera2014statistics}  & \yes & X & \yes &  &  &  &  &  &   \\ 
\cite{sokal1995biometry}  & [\yes]$^\text{e}$ & [D]$^\text{e}$ &  &  &  &  &  &  &   \\  \hline

\multicolumn{10}{l}{$^\text{a}$ By using the same equations, but without explicit mention.} \\
\multicolumn{10}{l}{$^\text{b}$ Only in the context of the one-sample problem.} \\
\multicolumn{10}{l}{$^\text{c}$ Only in the context of confidence intervals.} \\
\multicolumn{10}{l}{$^\text{d}$ Discusses how to handle zeros and ties, but does not mention them as consequences of continuity.} \\
\multicolumn{10}{l}{$^\text{e}$ Only in the context of regression or ANOVA.} \\
	\end{tabular}}
\end{table*}

Recall that $D=X-Y$ represents the per-topic difference between systems \textsf{X} and \textsf{Y}, and our hypothesis of interest is $H_0:\mu_D=0$. In other words, we test whether the expected difference between systems, in the topic population, is zero.
Under the assumption that $D$ is normally distributed with mean $\mu_D$ and variance $\sigma^2_D$, it follows from closure of the normal distribution that the sample mean $\overline{D}$ is also normally distributed with mean $\mu_D$ and variance $\sigma^2_D/n$, where $n$ is again the number of topics. The standardized mean, also known as the $z$-score, follows a standard normal:%\looseness=-1
\begin{equation}
	z=\frac{\overline{D}}{\sigma_D/\sqrt{n}} \sim\mathcal{N}(0,1)~.\label{eq:z}
\end{equation}
\noindent The $p$-value could be computed by placing this $z$-score in the \textsl{cdf} of the standard normal. Unfortunately, under typical experimental settings $\sigma_D$ is unknown. To address this, Student~\cite{student1908probable} introduced the $t$-score, replacing $\sigma_D$ with the \emph{observed} standard deviation $s_D$:
\begin{equation}
	t=\frac{\overline{D}}{s_D/\sqrt{n}} \sim\mathcal{T}(n-1)~,\label{eq:t}
\end{equation}
\noindent where $\mathcal{T}(n-1)$ is the $t$ distribution with $n-1$ degrees of freedom. Essentially, this accounts for the random noise in estimating $\sigma_D$ via $s_D$, making the $t$-distribution similar to a normal but with heavier tails. As $n$ increases though, it converges to a standard normal.

The main objection to the $t$-test arises when $D$ is non-normal---admittedly the norm in experimental research~\cite{micceri1989unicorn}---motivating the choice of the non-parametric Wilcoxon test. However, although the classical derivation of the test assumes the normality of $D$, we must note that this assumption has \emph{no practical relevance}. What really matters is whether the sample mean $\overline{D}$ remains normal, because the $p$-value is ultimately determined only by the $t$-score, regardless of the shape of $D$ itself. In this sense, the assumption of normality really concerns the mean, not the observed scores. Strictly speaking, Cram\'er's theorem tells us that $\overline{D}$ is exactly normal only if $D$ is normal as well, but exact normality is not the relevant criterion in practice. What matters is whether $\overline{D}$ is normal \emph{enough}. Indeed, by the Central Limit Theorem (CLT), $\overline{D}$ converges to a normal distribution as the sample size $n$ increases, provided that $D$ has finite non-zero variance. This means that the $t$-test is asymptotically correct under very general conditions, regardless of what distribution $D$ actually has~\cite[\S 13.2.1]{lehmann2022testing}. Therefore, the exact normality of $D$ is just a derivation condition for small sample problems; practical validity depends on the approximate normality of $\overline{D}$.

The left-hand side of Table~\ref{tab:t-wilcox} shows how the 25 Statistics textbooks describe the paired $t$-test. Only 11--16 books show the equivalence between the paired two-sample test of $\mu_X=\mu_Y$ and the one-sample test of $\mu_D=0$. This is important because, when the equivalence is omitted, readers may misinterpret the normality assumption as concerning $X$ and $Y$ when in reality it concerns $D$. This happens in 5 books, but it is perfectly possible for $X$ and $Y$ to diverge significantly from normality and yet result in a $D$ that is very close to normal. Another 9--13 books correctly placed it on $D$, and 6 books did not make it clear. Somewhat surprisingly, only 2 books identified the key assumption on the mean: \cite{box2005statistics} did this as a consequence of score normality, while only \cite{cetinkaya2024introduction} explicitly described $\overline{D}\sim\mathcal{N}$ as \emph{the} assumption. Another assumption easily misinterpreted is homoskedasticity (i.e., $\sigma_X^2=\sigma_Y^2$), which does not apply to our paired samples case.\footnote{Homoskedasticity would apply to the \emph{unpaired} two-sample case, for example when evaluating with two different collections~\citep{sakai2016two}.} This inconsistency in communication likely stems from the primarily didactic focus of textbooks: they usually explain the $t$-test through its formal derivation from normal variables, just as we did through Eqs.~\eqref{eq:z} and \eqref{eq:t}, while far less attention is given to the practical conditions under which the test remains valid when applied to real data.

\begin{figure*}[t]
\centering\includegraphics[scale=\figscale]{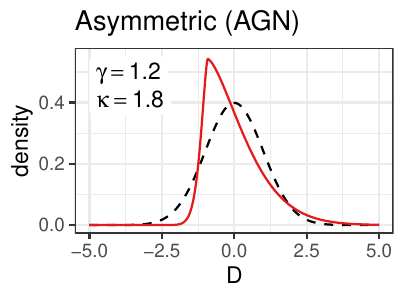}\includegraphics[scale=\figscale]{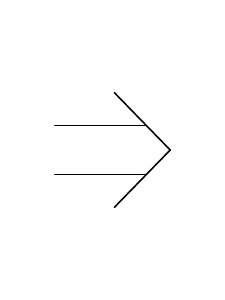}\includegraphics[scale=\figscale]{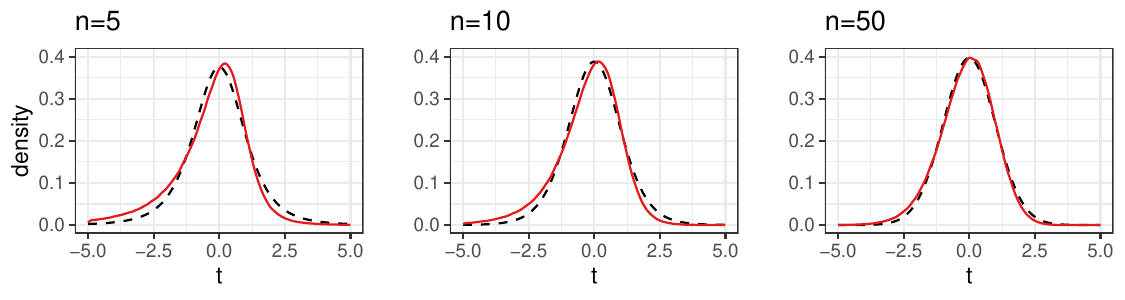}\\
\centering\includegraphics[scale=\figscale]{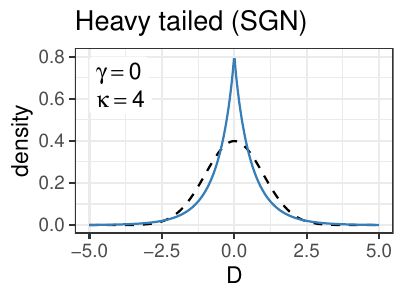}\includegraphics[scale=\figscale]{arrow.pdf}\includegraphics[scale=\figscale]{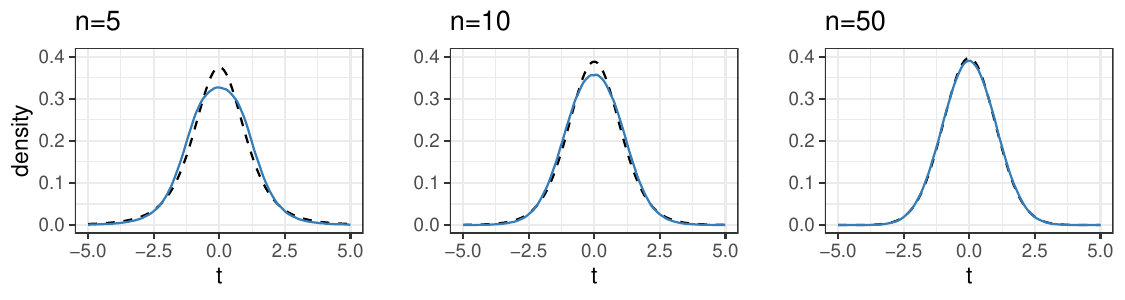}\\
\centering\includegraphics[scale=\figscale]{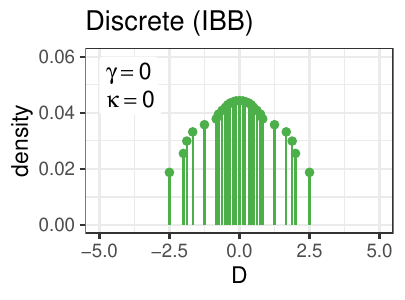}\includegraphics[scale=\figscale]{arrow.pdf}\includegraphics[scale=\figscale]{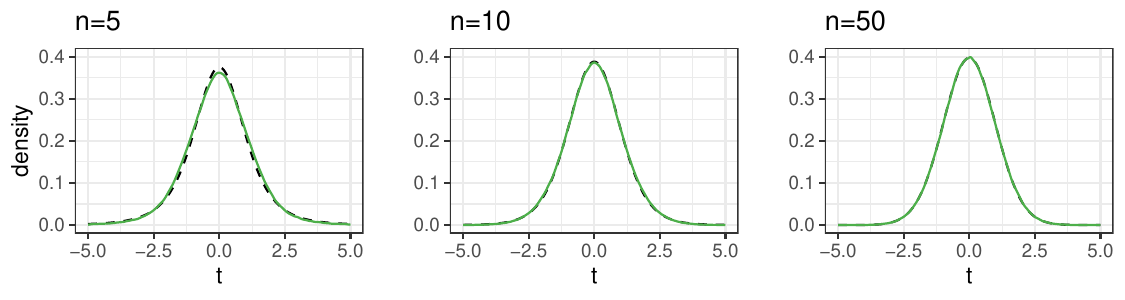}\\
\centering\includegraphics[scale=\figscale]{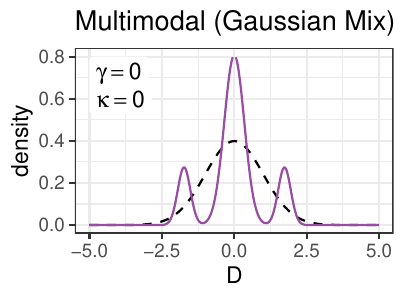}\includegraphics[scale=\figscale]{arrow.pdf}\includegraphics[scale=\figscale]{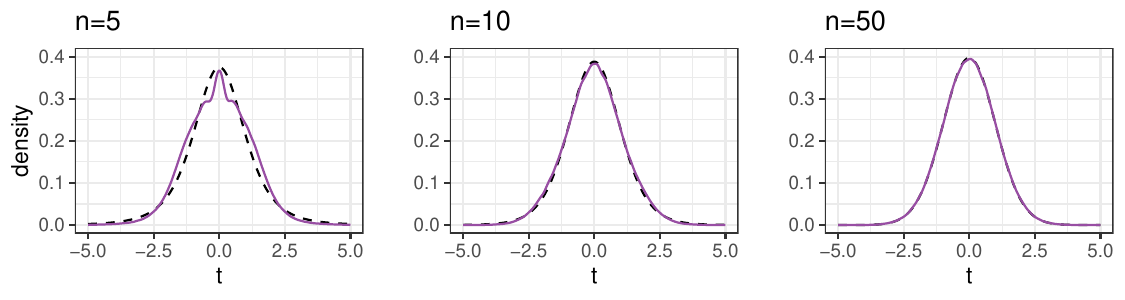}
%\centering\includegraphics[scale=\figscale]{clt-asymmetric_ref.pdf}\includegraphics[scale=\figscale]{clt-asymmetric_obs.pdf}\\
%\centering\includegraphics[scale=\figscale]{clt-asymmetric_ref.pdf}\includegraphics[scale=\figscale]{clt-asymmetric_obs.pdf}\\
%\centering\includegraphics[scale=\figscale]{clt-asymmetric_ref.pdf}\includegraphics[scale=\figscale]{clt-asymmetric_obs.pdf}\\
%\centering\includegraphics[scale=\figscale]{clt-asymmetric_ref.pdf}\includegraphics[scale=\figscale]{clt-asymmetric_obs.pdf}
\caption{Effect of asymmetry, tail heaviness, discrete support and multimodality of $\boldsec{D}$ (left), on the distribution of the $\boldsec{t}$ statistic for sample sizes $\boldsec{n=5,10,50}$ (right), illustrating the effect of the Central Limit Theorem. All distributions (see Appendix~\ref{app:distributions} for details) are standardized to zero mean and unit variance; $\boldsec{\gamma}$ and $\boldsec{\kappa}$ denote skewness and kurtosis. Sampling distributions are obtained empirically (1 million replicas). Dashed lines represent the distributions when $\boldsec{D}$ is normally distributed.}\label{fig:clt}
\end{figure*}

As noted earlier, the literature generally recommends a non-parametric alternative when normality is not present. As Table~\ref{tab:t-wilcox} shows, some books explicitly warn the reader about small $n$, skewness and heavy-tails.\footnote{We follow standard definitions, so skewness $\gamma$ refers to the third standardized moment and kurtosis $\kappa$ refers to the fourth. We always compute excess kurtosis.} To illustrate the effect of departures from normality directly in $D$, let us consider the four most common ways to break the assumption: a skewed distribution that introduces asymmetry, a high-kurtosis distribution that introduces heavy tails, a discrete and bounded distribution that modifies the support, even in irregular ways, and a multimodal distribution (see Figure~\ref{fig:clt}). These distributions are \emph{far} from normal, yet for moderate $n$ the empirical distributions of the resulting $t$-scores are nearly indistinguishable from the $\mathcal{T}(n-1)$ expected under normality. Therefore, in practice the $t$-test may behave well provided $n$ is not trivially small, even under gross violations of the normality assumption in $D$. Unfortunately, a similar discussion about normality and large samples is present only in 10 of the 25 books, needlessly pushing the reader towards the Wilcoxon test under the pretext of non-parametric safety.\looseness=-1

\subsection{The Misuse: Wilcoxon Signed Rank Test}\label{ssec:wilcox}
%%%%%%%%%%%%%%%%%%%%%%%%%%%%%%%%%%%%%%%%%%%%%%%%%%%%%%%%%%%%%%%%%%%%%%%%%%%%%%%%%%%%%%%%%%%%%%%%%%%%

Wilcoxon’s goal was not to create a test robust to non-normality, but to simplify calculations in the pre-computer era~\cite{wilcoxon1945individual}. His solution was rank-order statistics: instead of the raw differences $D_i$, the test works with their ranks $R_i$ after sorting by absolute value, thus discarding magnitudes. The test assumes \emph{exchangeability}, meaning that $X$ and $Y$ have the same distribution and the \textsf{X}/\textsf{Y} labels are arbitrary. Under $H_0$, a difference is equally likely to be positive or negative, so the statistic simply sums the ranks of positive differences:\footnote{For example, $D=\{-0.4,-0.1,0.4,0.8\}$ results in $R=\{2.5, 1, 2.5, 4\}$ and $W^+=6.5$.}\looseness=-1
\begin{equation}
	W^+=\sum_{i:~D_i>0}R_i~,
\end{equation}
\noindent yielding a $p$-value via the Wilcoxon signed-rank distribution with sample size $n$ (equivalently, one may sum the negative ranks $W^-$).

In his original formulation, Wilcoxon assumed $D$ to be continuous, as otherwise it would be impossible to tabulate the distribution of $W^+$ because it depends on the number of zeros and ties. Dropping zeros, as he proposed, remains the default in software packages, though alternatives exist~\cite{pratt1959remarks,conover1973methods}. When ties or zeros occur, a normal approximation is used for the distribution of $W^+$. Conover~\cite{conover1973rank} later showed that strict continuity is unnecessary; it suffices that $P(D=d)<1$ for all $d$, making the method applicable to ordinal data. Still, as Table~\ref{tab:t-wilcox} shows, 5 books continue to mention continuity as a requirement, while most of them simply focus on the practicalities of handling ties and zeros.  

But there is an additional subtlety with the Wilcoxon test: because it relies on rank-order statistics, it is formally a test for the \emph{median} of $D$, not the mean. This is explicitly mentioned in 11 books, mostly the ones on non-parametric methods. More importantly, a direct consequence of the exchangeability assumption is that the distribution of $D$ is \emph{symmetric} around 0, which makes the median equal to the mean. For this reason, the Wilcoxon test is legitimately used for the null hypothesis $H_0:\mu_D=0$, but at the price of the \emph{extra} assumption of symmetry. Unlike the $t$-test, whose validity improves with larger samples via the CLT, violations of symmetry are not diminished as $n$ increases. Quite the opposite! In the presence of asymmetry, a larger sample size only makes the systematic drift of $W^+$ even more pronounced. For this reason, the null hypothesis underlying the Wilcoxon test is effectively \emph{twofold}: it asserts both $\mu_D=0$ and symmetry of the distribution of $D$. If there is asymmetry, the null distribution of $W^+$ is no longer appropriate, and rejection may easily occur even when $\mu_D=0$. In such cases the test is no longer testing for a difference between groups, but for asymmetry in the distribution of differences.\footnote{Ironically, this is one of the uses of the Wilcoxon test statistic~\cite{li2014wilcoxon}.}

The contrast can be understood in terms of pivotality.\footnote{A statistic is called \emph{pivotal} if its sampling distribution does not depend on unknown parameters. A classical example is the $z$-statistic computed from a normal $\mathcal{N}(\mu,\sigma^2)$, as it follows a standard normal regardless of $\mu$ and $\sigma^2$.} The $t$-statistic is an exact pivot under normality and remains asymptotically pivotal under very mild conditions, as its limiting distribution does not depend on the shape of $D$. In contrast, the Wilcoxon $W^+$ statistic is distribution-free only under symmetry. When symmetry fails, its null distribution depends on the underlying distribution of $D$, and this dependence does not disappear as $n$ increases.

\subsection{Empirical Demonstration}
%%%%%%%%%%%%%%%%%%%%%%%%%%%%%%%%%%%%%%%%%%%%%%%%%%%%%%%%%%%%%%%%%%%%%%%%%%%%%%%%%%%%%%%%%%%%%%%%%%%%

To demonstrate the robustness of the $t$-test and Wilcoxon test to violations of their assumptions, as well as the effect of sample size, we carried out a small Monte Carlo simulation study of Type~I error rates under the null hypothesis $H_0:\mu_D=0$ and for each of the four distributions in Figure~\ref{fig:clt}. In particular, for each distribution we simulated 100K samples of sizes $n=5,50,500,5000$, and recorded the observed error rates at $\alpha=.05$. Recall that, under these conditions, the tests are expected to have a Type~I error rate of 5\%.
Table~\ref{tab:clt} shows that both tests are quite robust under the multimodal, heavy-tailed and discrete cases, although for Wilcoxon this holds only when the sample size is not extremely small. However, while the $t$-test is able to maintain the nominal 5\% error rate under asymmetry, the Wilcoxon test fails catastrophically, and increasingly so as the sample size increases. 

This simple demonstration shows that choosing the Wilcoxon test simply because of observing non-normal data can actually be a terrible idea: the symmetry assumption poses a much higher risk than non-normality does. This is ironic, to say the least, because asymmetry is probably the easiest feature to identify non-normality and turn to the Wilcoxon test, rendering it unreliable in the very situation for which it is most often recommended!

%%%%%%%%%%%%%%%%%%%%%%%%%%%%%%%%%%%%%%%%%%%%%%%%%%%%%%%%%%%%%%%%%%%%%%%%%%%%%%%%%%%%%%%%%%%%%%%%%%%%
\section{Impact on IR Research}\label{sec:impact}
%%%%%%%%%%%%%%%%%%%%%%%%%%%%%%%%%%%%%%%%%%%%%%%%%%%%%%%%%%%%%%%%%%%%%%%%%%%%%%%%%%%%%%%%%%%%%%%%%%%%

The previous section demonstrated that asymmetry \emph{can} severely distort the Wilcoxon test, but whether such distortions are likely in IR experimentation depends on how far IR data departs from normality. In this section, we therefore present simulations that progressively diverge from normality in controlled and interpretable ways, and evaluate both the $t$-test and Wilcoxon test with respect to their Type~I error rates. Our primary interest lies in the \emph{trends} as sample size increases and departures from normality become more severe, allowing us to draw general conclusions about robustness without tying them to specific configurations.

\subsection{Departures from Normality}\label{ssec:departures}
%%%%%%%%%%%%%%%%%%%%%%%%%%%%%%%%%%%%%%%%%%%%%%%%%%%%%%%%%%%%%%%%%%%%%%%%%%%%%%%%%%%%%%%%%%%%%%%%%%%%

\newcommand{\pcol}[1]{%
	\edef\val{\fpeval{#1}}%
	% red: < 0.025 or > 0.1
	\ifdim \val pt < 0.045pt
	\cellcolor{red!10}#1%
	\else\ifdim \val pt > 0.055pt
	\cellcolor{red!10}#1%
	% yellow: [0.025,0.045) or (0.055,0.1]
	\else\ifdim \val pt < 0.049pt
	\cellcolor{yellow!10}#1%
	\else\ifdim \val pt > 0.051pt
	\cellcolor{yellow!10}#1%
	% green: [0.045,0.055]
	\else
	\cellcolor{green!10}#1%
	\fi\fi\fi\fi
}

\begin{table}[t]
	\caption{Type~I error rates at $\boldsec{\alpha=.05}$ with the distributions from Figure~\ref{fig:clt} and various sample sizes. \textcolor{green!30}{$\blacksquare$} Green for good error rates (within .001 of $\boldsec{\alpha}$), \textcolor{yellow!30}{$\blacksquare$} yellow for reasonable rates (within .005 of $\boldsec{\alpha}$), and \textcolor{red!30}{$\blacksquare$} red for poor rates.}\label{tab:clt}
	\centering
	{\footnotesize\setlength{\tabcolsep}{1mm}\begin{tabular}{|r|rr|rr|rr|rr|}\hline
			~ & \multicolumn{2}{c|}{\textbf{Asymmetric}} & \multicolumn{2}{c|}{\textbf{Heavy tailed}} & \multicolumn{2}{c|}{\textbf{Discrete}} & \multicolumn{2}{c|}{\textbf{Multimodal}} \\ 
			\textbf{n} & \multicolumn{1}{c}{\textbf{$\boldsec{t}$-test}} & \multicolumn{1}{c|}{\textbf{Wlcxn}} & \multicolumn{1}{c}{\textbf{$\boldsec{t}$-test}} & \multicolumn{1}{c|}{\textbf{Wlcxn}} & \multicolumn{1}{c}{\textbf{$\boldsec{t}$-test}} & \multicolumn{1}{c|}{\textbf{Wlcxn}} & \multicolumn{1}{c}{\textbf{$\boldsec{t}$-test}} &  \multicolumn{1}{c|}{\textbf{Wlcxn}} \\ \hline
			5 & \pcol{.084} & \pcol{0}  & \pcol{.032} & \pcol{0}    & \pcol{.040} & \pcol{0}    & \pcol{.024} & \pcol{0} \\
			50 & \pcol{.056} & \pcol{.117} & \pcol{.048} & \pcol{.048} & \pcol{.050} & \pcol{.049} & \pcol{.051} & \pcol{.050} \\
			500 & \pcol{.052} & \pcol{.629} & \pcol{.051} & \pcol{.051} & \pcol{.051} & \pcol{.051} & \pcol{.050} & \pcol{.050} \\
			5000 & \pcol{.050} & \pcol{1} & \pcol{.050} & \pcol{.050} & \pcol{.050} & \pcol{.050} & \pcol{.050} & \pcol{.050} \\  \hline
	\end{tabular}}
\end{table}

Let us recall the four most common ways in which a distribution may depart from normality: asymmetry, tail heaviness, discrete support and multimodality. We must note that our goal is not to identify \emph{the} true distributional family underlying IR data,\footnote{One should question whether such an enterprise is even attainable in practice.} but rather to assess how progressively increasing departures along each of these dimensions affects test validity. We thus define 6 graded levels of departure from normality, labeled as low, medium, high, very high, extremely high, and pathologically high.

First, we deliberately do not assess the effect of multimodality because, unlike the other dimensions, even quantifying the amount of multimodality is itself a hard and ill-posed problem~\cite{silverman1981using,hartigan1985dip}: it does not naturally admit a scalar measure to be progressively increased, and both the number and prominence of modes depend on smoothing parameters, model assumptions and diagnostics. We therefore restrict attention to the more interpretable dimensions of asymmetry, tail heaviness and discreteness.

\begin{figure*}[t]
	\centering\includegraphics[scale=\figscale]{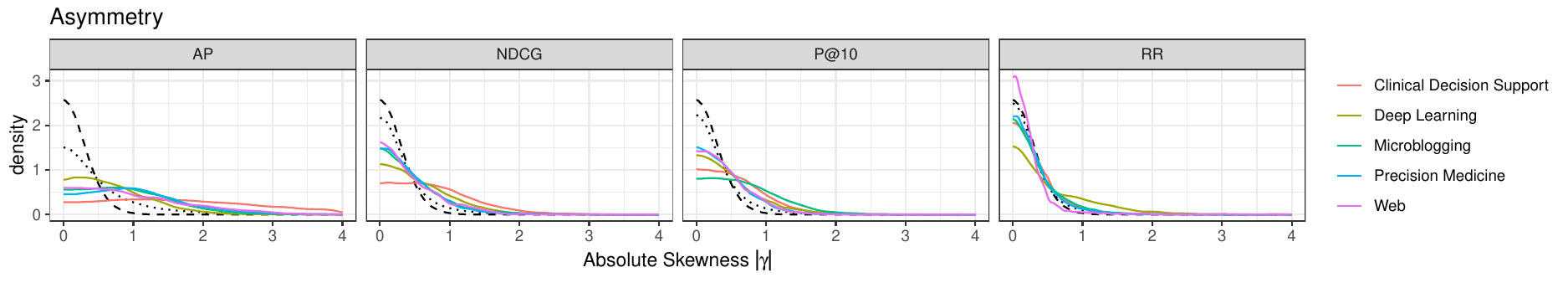}\\
	\centering\includegraphics[scale=\figscale]{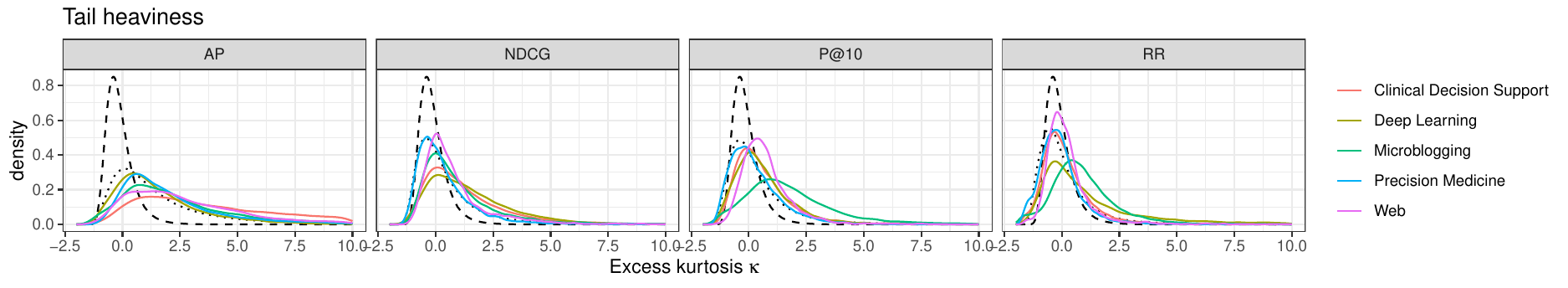}
	\caption{Symmetry and tail heaviness observed in $\boldsec{D}$ distributions from TREC data at $\boldsec{n=50}$. For reference, dashed lines represent the sampling distributions expected if $\boldsec{D}$ were normally distributed, and likewise dotted lines if they had tails as heavy as observed but still symmetric. The largest observed values are $\boldsec{|\gamma|\approx7}$ and $\boldsec{\kappa\approx45}$, but the axes are trimmed for clarity.}\label{fig:trec}
\end{figure*}

Regarding asymmetry and tail heaviness, we anchor the levels of departure in realistic regimes by first examining real IR metric scores from TREC. We collected data from somewhat recent and continued tracks with a large number of runs: Web (ad hoc) 2011--13, Microblog (ad hoc) 2012--14, Deep Learning (passage) 2019--21, Clinical Decision Support 2014--16 and Precision Medicine (abstracts) 2017--19. This resulted in 1,209 runs, and pairing all systems within the same track edition yielded 53,703 samples of paired differences $D$. For each pair, we computed AP, NDCG\footnote{For the Clinical Decision Support and Precision Medicine runs we followed the track guidelines and actually used infAP and infNDCG.}, P@10 and RR. 
Next, we computed skewness and excess kurtosis as measures of asymmetry and tail heaviness at $n=50$ topics (over/under-sampling where necessary). Figure~\ref{fig:trec} shows the distributions observed in TREC data and, for reference, the expected distributions under normality (dashed lines). It is clear that IR data depart substantially from the symmetry and tail heaviness expected from normally distributed scores. We note though that high sample skewness may still be observed with symmetric distributions if they have heavy tails, as we appear to have. Therefore, we also compare with the expected skewness under symmetric distributions with tails as heavy as observed in the data (dotted lines); it is clear that the observed skewness is still much higher than expected under symmetric populations.
To set departure levels for asymmetry, we use the values of (absolute) skewness that roughly correspond to the empirical percentiles $.25, .5, .75, .9, .99$ and $.999$: $\gamma=0.25, 0.5, 1, 1.5, 3, 5$, respectively (no departure would be $\gamma=0$). For tail heaviness we use excess kurtosis, treating heavier and lighter tails separately because their interpretation is not symmetric as in skewness (i.e. an excess kurtosis of $+0.5$ is not conceptually comparable to $-0.5$). For heavy tails, we again use values roughly corresponding to the same percentiles: $\kappa=0.5, 1.5, 3, 5, 15, 30$ (no departure would be $\kappa=0$). For light tails we cap calibration from below at $-1.2$, as more negative values correspond to U-shaped distributions rather than merely lighter-tailed shapes; modeling them would require explicit multimodal mechanisms, which we excluded from this study. The departure levels for light tails are $\kappa=-0.2, -0.4, -0.7, -0.9, -1.1, -1.2$. Together, these values define progressively increasing deviations from normality along each dimension.

Regarding discreteness, departures from normality arise naturally in IR metrics because their supports are finite and often irregular. This is particularly evident in metrics such as P@$k$ and RR@$k$ (e.g., a RR@$k$ score can only be one of $0,1/k,1/(k\!-\!1),...,1$).
We model discreteness via the support structure. For a metric M with cut-off $k$, let $\Omega$ denote the set of all possible values when computing the difference between two M@$k$ scores (rounded to 3 decimal digits for simplicity). The support of $D$ is $\Omega$, and the level of discreteness is governed by $k$ (e.g., $|\Omega|\!=21$ for P@10, whereas $|\Omega|\!=95$ for RR@10). To set departure levels for discreteness, we therefore use cut-off values $k=1000,500,100,50,10,5$. Larger $k$ produce fine-grained supports that approximate continuity, while smaller $k$ induce coarser supports. In the limit, $k\!\to\!\infty$ roughly corresponds to a continuous setting with no departure due to discreteness.\looseness=-1

\subsection{Non-Normal Distributional Mechanisms}\label{ssec:distributions}
%%%%%%%%%%%%%%%%%%%%%%%%%%%%%%%%%%%%%%%%%%%%%%%%%%%%%%%%%%%%%%%%%%%%%%%%%%%%%%%%%%%%%%%%%%%%%%%%%%%%

To generate distributions exhibiting the desired shapes, we employ the families described in Appendix~\ref{app:distributions}:
\begin{itemize}
	\item Asymmetry. We use the Asymmetric Generalized Normal (AGN) and Tukey G-and-H (TGH, with $h\!=\!0$) distributions, varying their asymmetry parameters $\xi$ and $g$ to match the target skewness while keeping tail heaviness controlled.
	\item Tail heaviness. We use the Symmetric Generalized Normal (SGN) and TGH (with $g\!=\!0$), varying their shape parameters $\beta$ and $h$ to match the target kurtosis while ensuring symmetry.\looseness=-1
	\item Discreteness. We use the Irregular Beta-Binomial (IBB), varying the support parameter $\Omega$ to match the target level of discreteness. Specifically, we simulate separately under P@k and RR@k regimes. Symmetry is guaranteed by construction, and tail heaviness is partially determined by the parameter $p$ to the underlying Beta-Binomial.
\end{itemize}
These distributions are chosen for flexibility, interpretability and control. We do not assess goodness-of-fit, and we of course do not claim that they represent the ``true'' data-generating mechanism underlying IR evaluation. Instead, we manipulate general distributional \emph{properties} ---symmetry, tail heaviness and support structure--- in a controlled manner. These families serve as useful parametric mechanisms to achieve target levels of these properties while holding others fixed, so we can focus on the impact of their related assumptions on Type~I error rates, not the adequacy of specific distributional models for IR data.
Finally, we hold $\sigma_D$ constant in all distributions, because Type~I error rates depend on distributional shape and sample size, not on scale. Rescaling does not affect skewness or kurtosis, cancels out in the $t$-statistic, and leaves Wilcoxon ranks unchanged. Since we already vary sample size $n$, allowing $\sigma_D$ to vary too would not change the trends of interest. Therefore, we fix $\sigma_D = 0.22$, corresponding to the median observed in TREC data.

\subsection{Type~I Error Rates under Non-Normality}
%%%%%%%%%%%%%%%%%%%%%%%%%%%%%%%%%%%%%%%%%%%%%%%%%%%%%%%%%%%%%%%%%%%%%%%%%%%%%%%%%%%%%%%%%%%%%%%%%%%%

For each distributional mechanism, departure level and sample size $n=5,50,500,5000$, we simulate 100K samples under $H_0:\mu_D=0$. We apply both the $t$-test and Wilcoxon test to every sample and record whether the null is rejected at $\alpha=.05$. These empirical rejection proportions provide estimates of the Type~I error rate in each condition. As emphasized earlier, our interest lies in the \emph{trend} of these rates as departures become more severe, rather than in the value observed at any single configuration.

Table~\ref{tab:ir-type1} reports the Type~I error rates. Starting at the bottom, for both discreteness and tail heaviness the $t$-test and the Wilcoxon test maintain error rates at the nominal $\alpha=.05$ across all practically relevant sample sizes. The only notable deviations occur at $n=5$. In that setting, the Wilcoxon test produces virtually no rejections across all configurations due to the limited number of attainable signed-rank scores, resulting in a conservative test. The $t$-test exhibits mild to large deviations under extreme tail behavior, yielding inflated error rates for light tails and deflated rates for heavy tails. However, these effects vanish rapidly as $n$ increases, and for $n \ge 50$ both tests remain well calibrated even under pathological tail conditions. Overall, departures from normality in the form of discreteness or tail heaviness do not meaningfully compromise Type~I error control in sample sizes typical of IR experimentation.

As expected, the situation changes drastically under asymmetry. For low and medium departures, the $t$-test maintains error rates at the nominal level or close. However, from very high departures onward, noticeable distortions emerge at smaller sample sizes. In particular, for the typical $n=50$ the $t$-test behaves just fine in the vast majority of cases, but it may inflate error rates up to approximately 10\% under pathological asymmetry. As sample size increases, the distortion diminishes due to the CLT, and for $n \ge 500$ the error rate returns to nominal levels even under extreme asymmetry.

In contrast, the Wilcoxon test fails \emph{systematically} under asymmetric distributions. At $n=5$ it again produces zero rejections due to discreteness constraints. For moderate sample sizes ($n=50$), it is close to nominal error rates only under low departures from symmetry. As asymmetry increases, the rejection rate rises very sharply. At $n=500$, even moderate asymmetry leads to substantial inflation, and under sufficiently strong asymmetry the test rejects the null hypothesis nearly always. This trend becomes more pronounced as sample size increases: rather than converging to nominal behavior like the $t$-test, the Wilcoxon test increasingly rejects under the null hypothesis of $\mu_D=0$ in the presence of asymmetry. As discussed earlier, this occurs because it eventually becomes a test of symmetry rather than a test of mean differences.

%%%%%%%%%%%%%%%%%%%%%%%%%%%%%%%%%%%%%%%%%%%%%%%%%%%%%%%%%%%%%%%%%%%%%%%%%%%%%%%%%%%%%%%%%%%%%%%%%%%%
\section{Perspectives}\label{sec:discussion}
%%%%%%%%%%%%%%%%%%%%%%%%%%%%%%%%%%%%%%%%%%%%%%%%%%%%%%%%%%%%%%%%%%%%%%%%%%%%%%%%%%%%%%%%%%%%%%%%%%%%

\subsection{On Test Optimality}
%%%%%%%%%%%%%%%%%%%%%%%%%%%%%%%%%%%%%%%%%%%%%%%%%%%%%%%%%%%%%%%%%%%%%%%%%%%%%%%%%%%%%%%%%%%%%%%%%%%%

The two most recent lines of work on statistical testing in IR studied test optimality via simulation, but framed the question in different ways. On the one hand, Urbano et al.~\cite{urbano2018stochastic,urbano2019statistical,urbano2021how} follow the orthodox perspective: one should choose the test that maximizes power \emph{subject to maintaining the nominal Type~I error rate}. A test unable to control its false positive rate is, under this view, simply invalid because it is testing a different hypothesis. On the other hand, Parapar et al.~\cite{parapar2020using,parapar2021testing} follow a trade-off perspective that emphasizes the compromise between Type~I errors and power: the most powerful test should be used to accelerate scientific innovation, \emph{even if it deviates mildly from the nominal Type~I error rate}.

Regardless of which of these perspectives one adopts, our results indicate that the Wilcoxon signed-rank test should not even be part of the discussion for IR experimentation. The reason is structural: Wilcoxon assumes symmetry of the distribution of paired differences. As Figure~\ref{fig:trec} shows, asymmetry is not an exotic edge case in IR data but rather the norm, and under such asymmetry it rejects the null hypothesis almost surely as sample size increases. In effect, it becomes a test of symmetry rather than a test of mean differences. Under the orthodox view, this alone is disqualifying. Under the trade-off view, the argument is even stronger: the Wilcoxon test cannot meaningfully trade false positives for power in the name of optimality, because its error rate is not even remotely close to the nominal level.\footnote{Works such as \cite{smucker2007comparison,urbano2019statistical} also concluded that the Wilcoxon test fails to maintain the nominal Type~I error rate, even under simulation regimes with little asymmetry \cite{urbano2021how}. In contrast, \cite{parapar2020using,parapar2021testing} reported that it does maintain the nominal level, which can perhaps be explained by how they simulate data under $H_0$: runs are independently generated from the same model and labeled \textsf{X} and \textsf{Y}, perfectly satisfying exchangeability.}

We emphasize that we do \emph{not} advocate the $t$-test as a universal solution. It appears here primarily as the natural counterpart to Wilcoxon, given that the latter is routinely proposed as its non-parametric alternative. Our central object of study is the Wilcoxon test and its behavior under empirically realistic departures from normality. Resampling-based procedures, such as permutation and bootstrap tests, have also been examined in the IR literature using both limited but real TREC data~\cite{smucker2007comparison,smucker2009agreement,urbano2013comparison}, as well as stochastic simulation of TREC-like data~\cite{urbano2018stochastic,urbano2019statistical,urbano2021how,parapar2020using,parapar2021testing}. These studies generally suggest that resampling tests maintain the nominal Type~I error rates in IR-like scenarios. However, they have not been tested as we did here, by analyzing how they are affected by departures from their key assumptions as seen in IR data. Extending the present analysis to these procedures is therefore a natural next step.

This connects to a broader methodological point. Attempting to determine \emph{the} ``true'' distribution of IR metric scores leads to an underdetermined problem: parametric, non-parametric, or relevance-based generative models all embed structural assumptions that cannot be validated beyond the observed samples. Our aim was precisely to avoid that goodness-of-fit rabbit hole by focusing instead on general distributional properties and their departure from test assumptions. The same principle applies when comparing inferential procedures: the question should not be which method aligns best with a specific configuration or generative model, but which procedures remain valid across realistic distributional regimes.

For similar reasons, we deliberately excluded multiple comparison procedures, because these corrections operate downstream of elementary tests and presuppose well-calibrated $p$-values. If a base test fails to control its Type~I error rate under realistic asymmetry, no multiplicity adjustment can repair that defect. What is clear is that the structural role of asymmetry should inform the assessment of any inferential framework, also under a Bayesian approach~\cite{carterette2015bayesian}.

\begin{table*}[t]
	\centering\caption{Type~I error rates at $\boldsec{\alpha\!=\!.05}$ under various levels of departure from normality with respect to asymmetry (top), tail heaviness (middle) and discreteness (bottom), as exhibited in IR-like data, and for various sample sizes. Color codes as in Table~\ref{tab:clt}.\looseness=-1}\label{tab:ir-type1}
	{\footnotesize\setlength{\tabcolsep}{1mm}\begin{tabular}{|r|rrrrrr|rrrrrr|} \hline
			& \multicolumn{6}{c|}{\textbf{$\boldsec{t}$-test}} 
			& \multicolumn{6}{c|}{\textbf{Wilcoxon}} \\
			\textbf{n}
			& \textbf{Low} & \textbf{Med.} & \textbf{High} & \textbf{Very H.} & \textbf{Extr. H.} & \textbf{Patho. H.}
			& \textbf{Low} & \textbf{Med.} & \textbf{High} & \textbf{Very H.} & \textbf{Extr. H.} & \textbf{Patho. H.}
			\\ \hline \hline
			
			& \multicolumn{12}{c|}{\textbf{Asymmetric}} \\ \cline{2-13}		
			& $\gamma=0.25$ & $\gamma=0.5$ & $\gamma=1$ & $\gamma=1.5$ & $\gamma=3$ & $\gamma=5$
			& $\gamma=0.25$ & $\gamma=0.5$ & $\gamma=1$ & $\gamma=1.5$ & $\gamma=3$ & $\gamma=5$ \\ \cline{2-13}	
			5 & \pcol{.052} & \pcol{.056} & \pcol{.070} & \pcol{.087} & \pcol{.136} & \pcol{.189} & \pcol{0} & \pcol{0} & \pcol{0} & \pcol{0} & \pcol{0} & \pcol{0} \\ 
			50 & \pcol{.051} & \pcol{.052} & \pcol{.054} & \pcol{.058} & \pcol{.075} & \pcol{.098} & \pcol{.052} & \pcol{.061} & \pcol{.092} & \pcol{.141} & \pcol{.316} & \pcol{.493} \\ 
			500 & \pcol{.051} & \pcol{.051} & \pcol{.051} & \pcol{.051} & \pcol{.054} & \pcol{.058} & \pcol{.077} & \pcol{.156} & \pcol{.452} & \pcol{.757} & \pcol{.992} & \pcol{1} \\ 
			5000 & \pcol{.049} & \pcol{.050} & \pcol{.050} & \pcol{.050} & \pcol{.051} & \pcol{.051} & \pcol{.321} & \pcol{.841} & \pcol{1} & \pcol{1} & \pcol{1} & \pcol{1} \\  \hline\hline
			
			& \multicolumn{12}{c|}{\textbf{Heavy tailed}} \\ \cline{2-13}	
			& $\kappa=.5$ & $\kappa=1.5$ & $\kappa=3$ & $\kappa=5$ & $\kappa=15$ & $\kappa=30$
			& $\kappa=.5$ & $\kappa=1.5$ & $\kappa=3$ & $\kappa=5$ & $\kappa=15$ & $\kappa=30$ \\ \cline{2-13}	
			5 & \pcol{.047} & \pcol{.043} & \pcol{.039} & \pcol{.036} & \pcol{.030} & \pcol{.028} & \pcol{0} & \pcol{0} & \pcol{0} & \pcol{0} & \pcol{0} & \pcol{0} \\ 
			50 & \pcol{.050} & \pcol{.050} & \pcol{.049} & \pcol{.049} & \pcol{.046} & \pcol{.045} & \pcol{.049} & \pcol{.049} & \pcol{.049} & \pcol{.049} & \pcol{.049} & \pcol{.049} \\ 
			500 & \pcol{.051} & \pcol{.051} & \pcol{.051} & \pcol{.050} & \pcol{.050} & \pcol{.049} & \pcol{.050} & \pcol{.050} & \pcol{.050} & \pcol{.050} & \pcol{.050} & \pcol{.050} \\ 
			5000 & \pcol{.050} & \pcol{.050} & \pcol{.050} & \pcol{.050} & \pcol{.050} & \pcol{.050} & \pcol{.050} & \pcol{.050} & \pcol{.050} & \pcol{.050} & \pcol{.050} & \pcol{.050} \\ \hline\hline
			
			& \multicolumn{12}{c|}{\textbf{Light tailed}} \\ \cline{2-13}	
			& $\kappa=-.2$ & $\kappa=-.4$ & $\kappa=-.7$ & $\kappa=-.9$ & $\kappa=-1.1$ & $\kappa=-1.2$
			& $\kappa=-.2$ & $\kappa=-.4$ & $\kappa=-.7$ & $\kappa=-.9$ & $\kappa=-1.1$ & $\kappa=-1.2$ \\ \cline{2-13}
			5 & \pcol{.052} & \pcol{.054} & \pcol{.057} & \pcol{.060} & \pcol{.064} & \pcol{.066} & \pcol{0} & \pcol{0} & \pcol{0} & \pcol{0} & \pcol{0} & \pcol{0} \\ 
			50 & \pcol{.051} & \pcol{.051} & \pcol{.051} & \pcol{.051} & \pcol{.051} & \pcol{.051} & \pcol{.049} & \pcol{.049} & \pcol{.049} & \pcol{.049} & \pcol{.049} & \pcol{.049} \\ 
			500 & \pcol{.051} & \pcol{.051} & \pcol{.050} & \pcol{.051} & \pcol{.050} & \pcol{.050} & \pcol{.050} & \pcol{.050} & \pcol{.050} & \pcol{.050} & \pcol{.050} & \pcol{.050} \\ 
			5000 & \pcol{.050} & \pcol{.050} & \pcol{.050} & \pcol{.050} & \pcol{.050} & \pcol{.050} & \pcol{.050} & \pcol{.050} & \pcol{.050} & \pcol{.050} & \pcol{.050} & \pcol{.050} \\ \hline\hline
			
			& \multicolumn{12}{c|}{\textbf{Discrete}} \\ \cline{2-13}	
			& $k=1000$ & $k=500$ & $k=100$ & $k=50$ & $k=10$ & $k=5$
			& $k=1000$ & $k=500$ & $k=100$ & $k=50$ & $k=10$ & $k=5$ \\ \cline{2-13}
			5 & \pcol{.043} & \pcol{.043} & \pcol{.042} & \pcol{.039} & \pcol{.037} & \pcol{.044} & \pcol{0} & \pcol{0} & \pcol{0} & \pcol{0} & \pcol{.001} & \pcol{.002} \\ 
			50 & \pcol{.050} & \pcol{.050} & \pcol{.050} & \pcol{.050} & \pcol{.049} & \pcol{.049} & \pcol{.049} & \pcol{.049} & \pcol{.049} & \pcol{.049} & \pcol{.048} & \pcol{.047} \\ 
			500 & \pcol{.050} & \pcol{.050} & \pcol{.050} & \pcol{.050} & \pcol{.050} & \pcol{.050} & \pcol{.050} & \pcol{.050} & \pcol{.050} & \pcol{.050} & \pcol{.050} & \pcol{.050} \\ 
			5000 & \pcol{.050} & \pcol{.050} & \pcol{.050} & \pcol{.050} & \pcol{.050} & \pcol{.050} & \pcol{.051} & \pcol{.050} & \pcol{.050} & \pcol{.050} & \pcol{.050} & \pcol{.050} \\ \hline
	\end{tabular}}
\end{table*}

\subsection{On Importing Results from Other Fields}
%%%%%%%%%%%%%%%%%%%%%%%%%%%%%%%%%%%%%%%%%%%%%%%%%%%%%%%%%%%%%%%%%%%%%%%%%%%%%%%%%%%%%%%%%%%%%%%%%%%%

A broader perspective concerns the way methodological results travel across disciplines. The review exercise in Section~\ref{sec:stats-lit} illustrates a twofold risk. First, isolated results from another field may not reflect the full scope of debate within that field. In our case, the Statistics literature contains extensive discussions on robustness, asymptotics, and assumption violations (e.g., \cite{chaffin1993effect,cicchitelli1989robusness,sawilowsky1992more,box1953non,boneau1960effects,lumley2002importance,blair1985comparison}), which are rarely captured by general summaries. Second, even canonical sources like textbooks may present views that are historically contingent or internally contested. Therefore, we should avoid methodological imports from other fields based on single sources, as it risks the authoritative introduction of knowledge that may itself be contested or even outdated. We need broader scrutiny, as otherwise we may inadvertently hard-code contested views into IR practice and even into calls for papers or reviewing guidelines~\cite{sakai2020fuhr}.

In our case, and setting aside the larger discussion around definitions, the Wilcoxon test is frequently introduced in textbooks simply as the non-parametric alternative to the $t$-test, sometimes described as appropriate only for ordinal data. Such statements, when taken in isolation, obscure the nuance required to interpret Wilcoxon as a test of central tendency. For instance, if one relied on only one of the books in Table~\ref{tab:t-wilcox}, there would be roughly a 50\% chance of missing the symmetry assumption; indeed, only about half of the IR papers cited in Section~\ref{sec:intro} mention it. The issue, however, is not merely that the assumption exists, but that its origin, implications and practical consequences for IR evaluation have remained insufficiently clarified. We fill that gap in this paper.
%Although symmetry is formally a weaker assumption than normality, violating the former has a much larger impact on the Wilcoxon test than violating the latter has on the $t$-test.

A parallel example in IR is the recent debate on levels of measurement: arguments based on Stevens’ taxonomy have been used to question averaging certain metrics~\cite{fuhr2018some,sakai2020fuhr,ferrante2021towards,moffat2022batch}, in spite of the broad discussion and criticism around this taxonomy, and the existence of others~\cite{michell1986measurement}. As pointed out in Section~\ref{ssec:recs}, the Wilcoxon test has itself been subject to similar debates over the years.

\subsection{On Moving from Testing to Modeling}
%%%%%%%%%%%%%%%%%%%%%%%%%%%%%%%%%%%%%%%%%%%%%%%%%%%%%%%%%%%%%%%%%%%%%%%%%%%%%%%%%%%%%%%%%%%%%%%%%%%%

A broader methodological issue underlies the discussion around statistical tests: the reliance on contrast-based analyses that isolate a single experimental factor and treat all others as fixed or non-existent. This paradigm appears in classical pairwise comparisons between systems and, increasingly, in ablation studies that toggle individual system components and compare specific variations. Such approaches allow us to assess whether a change produces a statistically detectable difference, but they account for only one source of variability (i.e., topics). They do not allow us to quantify how much variability is attributable to other factors, control for them when comparing systems, or reveal how components interact across datasets or conditions.

Several lines of work already point toward a more explicit modeling approach. For example, Bootstrap ANOVA and related techniques have been proposed to account for document collection variation and system--collection interactions, allowing uncertainty to reflect not only topic sampling but also corpus variability~\cite{voorhees2017using,zobel2020corpus,ferro2019using}. Earlier, Generalizability Theory was applied to IR test collections to decompose variance across facets such as topics and assessors \cite{bodoff2008test}. While limited in flexibility, this line of work made it clear that topic sampling is only one of many components of experiment uncertainty \cite{taguesutcliffe1992pragmatics,canamares2020offline}.
More recently, linear and generalized linear modeling approaches have been explored to test system differences and to study the contribution of system components within unified regression frameworks \cite{ferro2016general,faggioli2022detecting,carterette2012multiple}. These approaches allow us to control for multiple factors other than topics, handle interaction effects, increase statistical power and use structured error components, thereby moving beyond repeated pairwise contrasts~\cite{harrell2015regression}.\looseness=-1

The perspective we advocate for is therefore not merely to replace one test with another, but to move from isolated hypothesis tests toward explicit analysis models that make assumptions transparent, allow diagnostics, and properly handle multiple sources of variance in a single framework. In that setting, significance tests become downstream summaries of a structured model rather than standalone decisions applied to aggregated topic differences.

%%%%%%%%%%%%%%%%%%%%%%%%%%%%%%%%%%%%%%%%%%%%%%%%%%%%%%%%%%%%%%%%%%%%%%%%%%%%%%%%%%%%%%%%%%%%%%%%%%%%
\section{Conclusions}
%%%%%%%%%%%%%%%%%%%%%%%%%%%%%%%%%%%%%%%%%%%%%%%%%%%%%%%%%%%%%%%%%%%%%%%%%%%%%%%%%%%%%%%%%%%%%%%%%%%%

We have revisited the role of statistical testing in IR evaluation, not by adding yet another empirical study but by systematically examining the foundations that shape our practice: Statistics textbooks. Our review shows that they routinely present a simplistic parametric vs. non-parametric dichotomy, reinforcing the idea that the Wilcoxon test is the ``safe'' alternative to the $t$-test whenever normality is in doubt. We find this narrative misleading and dangerous.\looseness=-1

We identified three core problems. The \emph{myth} is that non-parametric methods are assumption-free, when in fact they have assumptions of their own; Wilcoxon silently demands symmetry. The \emph{misconception} is that the $t$-test requires normal data, when it really relies on the approximate normality of the sample mean, a condition satisfied in practice through the Central Limit Theorem. The \emph{misuse} is the routine application of Wilcoxon instead of the $t$-test while ignoring symmetry, which leads to distorted null distributions and inflated error rates even under the null hypothesis. As our test collections continue to grow, these issues do not fade away but are amplified: Wilcoxon becomes even more sensitive to slight departures from symmetry, virtually guaranteeing a rejection of the null, and probably for the wrong reason.

The implications for IR are severe. Simulations confirm that Wilcoxon fails precisely under the conditions that dominate IR data, while the $t$-test remains robust. What is widely seen as the conservative and safe choice is, in fact, the one carrying higher risk. The Wilcoxon test should be retired from IR evaluation.

\appendix

\section{Non-Normal Distributions}\label{app:distributions}

This appendix describes the distribution families we used to generate non-normal data; some were illustrated already in Figure~\ref{fig:clt}:%\looseness=-1
\begin{itemize}
	\item Symmetric Generalized Normal (SGN)~\cite{subbotin1923law,nadarajah2005generalized}. This is a generalization of the normal distribution where parameter $\beta$ controls tail heaviness. It is symmetric by design.
	\item Asymmetric Generalized Normal (AGN)~\cite{fernandez1998bayesian}. This is another generalization of the normal distribution where parameter $\xi$ controls asymmetry and parameter $\nu$ controls tail heaviness.
	\item Tukey G-and-H (TGH)~\cite{tukey1977modern,headrick2008parametric}. This family is defined via transformations of a normal variable, where parameter $g$ controls asymmetry and parameter $h$ controls tail heaviness.
%	\item Symmetric Generalized Normal (SGN)~\cite{subbotin1923law,nadarajah2005generalized}, with parameters $\mu$ for location, $\sigma$ for scale, and $\beta$ for shape. This is a generalization of the normal distribution where parameter $\beta$ allows for heavier or lighter tails. As the name suggests, it is symmetric by design.
%	\item Asymmetric Generalized Normal (AGN)~\cite{fernandez1998bayesian}, with parameters $\mu$ for location, $\sigma$ for scale, $\xi$ for asymmetry and $\nu$ for shape. This is another generalization of the normal distribution where parameter $\xi$ controls asymmetry and parameter $\nu$ controls tail heaviness.
%	\item Tukey G-and-H (TGH)~\cite{tukey1977modern,headrick2008parametric}, with parameters $\mu$ for location, $\sigma$ for scale, $g$ for skewness, and $h$ for tail heaviness. This family is defined via transformations of a normal variable, where parameter $g$ allows for asymmetry and parameter $h$ allows for heavier or lighter tails.
\end{itemize}
These distribution families allow us to vary skewness and kurtosis while offering different tail decay behaviors.

For the generation of discrete data as if produced by an IR metric M, we should consider the set of all possible values when calculating the difference between two scores; let us refer to this set as $\Omega$.
Our goal is therefore to simulate discrete data with an arbitrary support $\Omega$. To do so, we use a Beta-Binomial distribution with parameters $n=|\Omega|-1$ and $\alpha=\beta=p$ to generate an integer random variable $J$, which can then be used to index $\Omega$. We call this the Irregular Beta-Binomial (IBB). Formally, $D\sim IBB(\Omega,p)$ is calculated as $D=\Omega_{J-1}$, where $J\sim BetaBin(|\Omega|-1,p,p)$. By construction, IBB is therefore symmetric and tail heaviness depends on $\Omega$ and $p$. Note that the amount of ``discreteness'' (i.e., the size of $\Omega$ and the spacing between its values) is entirely determined by M.

\begin{acks}
This is dedicated to those who, somehow, always thought there was something off with FC Barcelona; to those who finally think so now that the Negreira case is public; to those who think, deep down, that nothing will happen or change; and to Pepe Kollins for ensuring we all keep thinking about it. Keep thinking.
\end{acks}

\bibliographystyle{ACM-Reference-Format}
\balance
\bibliography{sigir2026}

\end{document}